\documentstyle[12pt,aps,epsfig]{revtex}

\def\beq{\begin{equation}} 
\def\eeq{\end{equation}}
\def\ber{\begin{eqnarray}} 
\def\eer{\end{eqnarray}}
\def\p{{\rm I\!P}} 
\def\C{{\cal C}}
\def\ket#1{| #1 \rangle}

\def\ci#1{$^{(\ref{#1})}$}
\def\cii#1#2{$^{(\ref{#1},\ref{#2})}$}
 
\begin{document}

\title{{\hfill IUCAA-38/97} \\
ORIGIN OF QUANTUM RANDOMNESS \\ IN THE PILOT WAVE QUANTUM MECHANICS}
\author{Yu.~V.~Shtanov}  
\address{Inter-University Centre for Astronomy and Astrophysics, \\
Post Bag 4, Ganeshkhind, Pune 411 007, India  \\ and  \\ 
Bogoliubov Institute for Theoretical Physics,  
Kiev 252143, Ukraine\footnote{Permanent address. Electronic mail: 
shtanov@ap3.gluk.apc.org}} 
\date{May 13, 1997} 
 
\maketitle \bigskip

\bigskip 
 
\begin{abstract} 
We account for the origin of the laws of quantum probabilities 
in the de~Broglie-Bohm (pilot wave) formulation of quantum theory by  
considering the property of ergodicity likely to characterise the dynamics 
of microscopic quantum systems.  
\end{abstract}

    \sloppy                           
 
\section{Outlook of the theory. The problem} 
 
 Pilot wave quantum mechanics, the theory put forward by Bohm,\ci{Bohm}  
and based on the earlier ideas of de~Broglie,\ci{de Broglie} was  
aimed at resolving the notorious ``measurement problem'' inherent in quantum  
theory. The source of this problem can be seen in the twofold meaning 
attributed to the   
wave functions that describe the quantum experiment. 
Wave functions that  
we ascribe to microscopic invisible quantum systems (such as elementary  
particles) are eventually used to represent {\em probability amplitudes} for  
the observable events that take place with our measuring devices.   
Wave functions ascribed to the devices themselves receive a vague definition 
of representing their {\em states}, that, in turn, must correspond to  
observers' sensible impressions of the devices. In such a twofold definition  
of the basic quantity lies the source of difficulties and paradoxes of  
various kinds, the famous Schr\"{o}dinger cat paradox being one of them.  
One of the major merits of the pilot wave formulation is that it  
overcomes this situation.  
 
 The basic idea of the pilot wave theory is the following.  
Every closed physical system is described  
by a deterministic evolution of configuration variables (which  
Bell\ci{Bell} has called ``{\em be}ables''). These are 
the same as in the classical physics and are just the spatial coordinates 
of the elementary particles and the configurations of various fields. 
The evolution of the 
configuration variables is guided (piloted, in de~Broglie's terminology)  
by a quantum wave that obeys the Schr\"{o}dinger equation. 
The probabilistic character of this mechanics is merely the consequence 
of our (essential) ignorance of and inability to control the actual values  
of particle and field microscopic configuration variables. 

Besides nonrelativistic quantum mechanics the  
pilot wave interpretation has been also applied to the relativistic 
theory of particles and bosonic (scalar and vector) fields, and was argued 
to be consistent with the observable special relativity.\cii{BH}{Holland}\  
A straightforward  
extension to quantum geometrodynamics was made by Holland\ci{Holland} 
and by Horiguchi\ci{Horiguchi} and  
further studied by the author.\ci{Shtanov}\ A complete pilot wave quantum  
theory of particles and fields (sketched in Ref.~\ref{BH}, see also  
Ref.~\ref{Shtanov}) still remains to be developed.    

 Consider the pilot wave  
theory in more detail in nonrelativistic quantum mechanics. A set of $N$  
nonrelativistic spinless particles are described by their spatial coordinates  
${\bf x} \equiv \left({\bf x}_1, \ldots,  {\bf x}_N\right)$. 
The wave function  
$\psi$ of such a system obeys the Schr\"{o}dinger equation  
\beq  
i \hbar {\partial \psi \over  
\partial t} = - {\hbar^2 \over 2} \sum_n {1 \over m_n} \triangle_n \psi + V  
\psi \, , \label{schrod}  
\eeq  
where $V = V({\bf x})$ is the particle  
interaction potential. If one represents the wave function in the polar form  
as $\psi = R \exp \left( i S / \hbar \right)$ then from Eq.~(\ref{schrod}) it  
follows that the phase $S({\bf x},  t)$ and the amplitude $R({\bf x},   
t)$ satisfy the system  
\ber  
{\partial S \over \partial t} + \sum_n {1 \over  
2m_n} \left(\nabla_n S \right)^2 + V + Q &=& 0 \, , \label{qhj} \\ 
{\partial R^2 \over \partial t} + \sum_n {1 \over m_n} \nabla_n \left(R^2 
\nabla_n S \right) &=& 0 \, , 
\eer 
where 
\beq 
Q = - \sum_n {\hbar^2 \over 2m_n} {\triangle_n R \over R} \label{qpot} 
\eeq 
is the so-called quantum potential. 
In the pilot wave formulation of quantum mechanics the evolution of 
the coordinates ${\bf X}$, that correspond to the arguments ${\bf x}$ of the 
wave function, is governed by $\psi({\bf x},t)$ via the guidance equation 
\beq 
m_n \dot {\bf X}_n = {i\hbar \over 2}\left({\psi\nabla_n\psi^* - 
\psi^*\nabla_n\psi \over |\psi|^2}\right)_{{\bf x} = {\bf X}} = 
\nabla_n S ({\bf X},t) \, . \label{guidcon} 
\eeq 
The equation (\ref{qhj}) for $S({\bf x},t)$ is the quantum generalisation 
of the classical 
Hamilton-Jacobi equation and differs from the latter only by the presence 
of the quantum potential $Q({\bf x},t)$. 
The guidance equation (\ref{guidcon}) is expressed in terms of the function
$S({\bf x},t)$ in the same form as in the classical theory it is expressed 
in terms of 
the solution to the Hamilton-Jacobi equation. Hence, in the limit in which 
the quantum potential $Q$ in (\ref{qhj}) can be 
neglected we recover classical evolution. We thus see that new formulation 
of quantum theory can be regarded as just a ``deformation'' of the classical 
dynamics (general discussion of this analogy with the classical case can be  
found in Ref.~\ref{Holland}). Note that in the present interpretation the temporal  
dynamics of the particle coordinates completely determines  
the physical state of a system, be it microscopic or macroscopic.  
The role of the wave function in every physical situation is one  
and the same, namely, to provide the guidance laws for configuration 
variables. Thus the description of the physical systems becomes unified, and  
the abovementioned source of the difficulties, which existed in the twofold  
character of such a description, is eliminated. 

 The formalism of quantum dynamics outlined above can be readily applied  
to the case of a single closed quantum system. In practice, however, we  
usually deal with what we call quantum ensembles that are collections of many 
identical systems each piloted in the way described above.  
If all these systems are piloted by one and the same  
wave function then an ensemble is called pure. Otherwise it is called mixed.  
(Note that the systems that form such ensembles need not exist  
simultaneously: experiments with one and the same arrangement may be  
carried out repeatedly.) 
In the pilot wave formulation of quantum mechanics the measurement process is  
regarded as just a partial case of the generic evolution guided by a  
wave function that obeys the Schr\"{o}dinger equation.  
The probabilistic character of measurement outcomes is caused by 
our ignorance of and inability to control the actual  
(initial) values of particle and field microscopic configuration variables in  
each system of an ensemble as well as in the measuring apparatus.  

 Consider general description of an ideal measurement of an observable 
$\Lambda$, with a discrete spectrum $\Lambda_n$ and the corresponding 
normalised eigenstates 
$\xi_n \left(x_S\right)$, of a system described by a set of coordinates 
$x_S$. Let the measuring apparatus be described by a set of variables $x_A$
and let its initial wave function be $\phi\left(x_A\right)$. The initial wave 
function of the total system is 
\beq
\psi_i \left(x_S, x_A\right) = \xi\left(x_S\right)\phi\left(x_A\right) = 
\sum_n c_n \xi_n\left(x_S\right)\phi\left(x_A\right)\, . \label{init}
\eeq
Suppose that due to interaction between the system and the measuring apparatus
the wave function evolves into
\beq
\psi_f\left(x_S, x_A\right) = \sum_n c_n \xi_n\left(x_S\right)\phi_n
\left(x_A\right)\, , \label{fin}
\eeq
with nonoverlapping, macroscopically distinct, normalised states $\phi_n$ of 
the 
measuring apparatus. In the course of the measurement process the corresponding
configuration variables $X_S$ and $X_A$ evolve in a definite way depending 
on their initial values, and at the end of the experiment the variables 
$X_A$ turn out to be in a localisation region of one, and only one, of the 
states $\phi_n$. Then the macroscopic state of the apparatus, hence, the 
measurement outcome, is uniquely specified in an experiment over a particular 
system. 

 If we have an ensemble of measurements described by Eqs.~(\ref{init}) and 
(\ref{fin}) then the outcomes will be random due to random distribution
of the initial variables $X_S$ and $X_A$. In the pilot wave mechanics, 
in order that the probabilities $p_n$ of different measurement outcomes 
coincide with those of the standard (Copenhagen) approach, $p_n = \left|
c_n \right|^2$, it is necessary to assume that the  
configuration variables of the systems in a quantum ensemble are  
distributed in accord with their wave function, so that $p(x) = |\psi(x)|^2$,  
where $x$ denotes the set of all configuration variables, and $p(x)$ is their  
distribution function. Such a condition is sometimes called  
{\em quantum equilibrium}\cii{Valentini}{DGZ}: it is a consequence of the 
Schr\"{o}dinger  
equation that provided the equality holds initially for a given ensemble, 
it will hold at all times (so long as the ensemble remains closed).  
However, in the framework of the theory discussed, one has to explain the   
origin of such a distribution. This question, which is crucial for the pilot  
wave interpretational scheme, will be the focus of the discussion  
in this paper.

\section{Attempts at a solution} 
 
 Bohm himself\cii{Bohm1}{BH} gave qualitative reasoning with regard to 
this problem. The (typically) complicated, quasirandom, motions of 
interacting particles, he argued, would lead to the establishment of 
quantum equilibrium. If one defines function $f(x)$ by $p(x) = 
f(x) |\psi(x)|^2$, then it is easy to see that $f(x)$ is conserved along the 
trajectories, and the conjecture made by Bohm was that due to the 
complicated character of these trajectories the {\em coarse-grained} value of 
$f(x)$ will approach unity, thus $p(x)$ will approach $|\psi(x)|^2$ as 
{\em coarse-grained} values, what may be sufficient for all practical 
purposes.  However, 
he has not succeeded in justifying these insights quantitatively. Perhaps 
for this reason in the modified pilot wave proposal of Bohm and Vigier\ci{BV} 
(see also Ref.~\ref{BH}) 
an additional external stochastic force was added to the right-hand  
side of the guidance equations (\ref{guidcon}) in order to account for  
the occurrence of quantum equilibrium. A similar theory was also put 
forward by Nelson.\ci{Nelson}\ In this paper we consider only the original  
``minimal'' version of the pilot wave theory as it is expressed in  
Ref.~\ref{Bohm}.  
 
 Among the recent approaches to the problem of quantum equilibrium that  
we are aware of, one is due to Valentini\ci{Valentini} 
and another is due to D\"{u}rr, Goldstein and Zangh\`{\i}.\ci{DGZ}\   
While the approach of Valentini can be regarded as an elaboration of  
Bohm's argument (see above), that of D\"{u}rr {\em  et al.\@} is based on  
quite a different idea.  
To our mind, however, the proofs and demonstrations contained in  
Ref.~\ref{Valentini} and in Ref.~\ref{DGZ}  
do not achieve the goal, for the reasons that follow.  
 
 Valentini\ci{Valentini} made an attempt to justify the Bohm's conjecture 
that for a pure ensemble of closed {\em complicated} systems the  
coarse-grained distribution $\overline p(x)$ of the configuration  
variables will approach the  
coarse-grained value $\overline{|\psi(x)|^2}$ (here overline  
denotes coarse graining). The corresponding analysis involves 
the quantity $S = - \int \overline p \log\left(\overline p /  
\overline{|\psi|^2}\right)dx$ called ``subquantum entropy.'' By analogy  
with the classical statistical mechanics (reasoning based on Boltzmann's  
$H$-theorem) it is suggested by the author\ci{Valentini} that this quantity will  
increase in time approaching its maximum value of zero, thereby leading to  
coarse-grained quantum equilibrium, $\overline p = \overline{|\psi|^2}$. 
Such a suggestion is based solely on the  
fact (called ``subquantum $H$-theorem'' by Valentini\ci{Valentini}) that if the  
conditions $p = \overline p$ and $|\psi|^2 = \overline{|\psi|^2}$ (the  
conditions of ``no fine-grained microstructure,'' assumed to hold at the  
initial moment of time) are valid, then the above-presented coarse-grained  
``entropy'' $S$ acquires its local minimum at that moment of time. This  
property of $S$, however, does not seem to be sufficient for the conjecture  
to be justified, if only because it will hold equally well for systems that  
will never approach quantum equilibrium (see examples in the final part of  
this paper).  
                                    
 Demonstration of D\"{u}rr {\em et~al.}\ci{DGZ} is based on the  
notion of {\em typicality} which is applied to the domain of all possible  
initial conditions of a model universe. Specifically, the modulus squared 
$\left|\Psi\right|^2$ of the universal wave function is taken to represent  
the measure density of typicality  
in the domain of configuration variables. This 
measure is singled out on the basis of its equivariance, which means  
that at any moment of time it is expressed through $\Psi$ in one and the same  
manner. The authors then show that the  
set of initial conditions that would conform (to certain precision) with  
the usual quantum mechanical statistical predictions has measure of  
typicality close to one. To our mind, equivariance of the 
{\em specific subjective} measure introduced, although important 
property, is not sufficient for regarding this measure as   
relevant to {\em objective} distributions encountered in the  
experiments. Especially as it was noted by the authors  
themselves (Sec.~7 of Ref.~\ref{DGZ}) that a different choice of the measure  
for typicality would result in predicted probability distributions 
different from the observed ones.

\section{Ergodicity argument for stationary states} 
 
 In this paper we suggest that arguments of the {\em ergodic theory}    
can be used to justify the quantum equilibrium hypothesis.  
An approach of such a kind has been noted by Valentini 
(see p.~40 of Ref.~\ref{Valentini}(b)) but rejected in view of one of the 
difficulties inherent in it, what will be discussed below. 
 
 Let us take the point of view, adopted in the classical statistical 
mechanics (see, e.g., Landau and Lifshits\ci{LL}), that equilibrium ensemble 
averages of  
various functions of dynamical variables can be represented by their time  
averages. The theory developed along this line of reasoning is the 
ergodic theory (for  
an introduction to which see Ref.~\ref{HWSLM}). A dynamical system in this  
theory is regarded as a measure space together with one-parameter  
(discrete or continuous) group of measure-preserving transformations.
A subset in the space of dynamical variables is called {\em invariant set} 
if it is invariant (modulo set of measure zero) 
with respect to all these transformations.
A dynamical system is called {\em ergodic} if for any of its invariant sets the 
measure either of this set, or of its complement, is zero. 
As a consequence of the 
Birkhoff-Khinchin  ergodic theorem, the fraction of time  
spent by an ergodic system in a measurable region $\Omega$ of its 
dynamical variables  
tends to a value proportional to the invariant measure of this region as time  
goes to infinity. For example, in the case of the Hamiltonian dynamics such 
an invariant  
measure is the surface measure, induced by the Liouville measure, on the  
constant energy surface in the phase space.  Justification of the 
microcanonical  
equilibrium distribution then reduces to the proof (which is usually a  
difficult task) or assumption of ergodicity of a particular system.  Note,  
that the ergodicity property can be formulated in terms of {\em any} measure  
equivalent\footnote{Two measures with common domain are said to be 
{\em equivalent}  
if they have as mutual all the sets of measure zero.} to the invariant  
measure, in this sense ergodicity does not rely strongly on this latter.  
On the other hand, for an ergodic system the invariant measure is unique in  
the corresponding equivalence class.  
 
 If we take all this to be of equal significance to the pilot wave quantum 
mechanics, we can relate the distribution of the configuration variables 
to the measure density $\left|\Psi\right|^2$ using the ergodicity property 
of the corresponding pilot wave dynamic flow. We must consider several 
problems with this approach. The fact that the measure with density 
$\left| \Psi \right|^2$ is in general time-dependent, hence only equivariant 
rather than also invariant, calls for an essential modification of the 
above argument,  
as compared to the classical case [it is this difficulty that has been 
noted on p.~40 of Ref.~\ref{Valentini}(b)]. This problem will be simply
avoided if one restricts attention to systems in stationary states. 
This is what we shall do first. Incidentally, this is just  
what takes place when one proceeds to the {\em universal} level (as suggested  
by D\"urr {\it et al.}\ci{DGZ}) and takes into account {\em general covariance} of the  
complete theory that includes gravity. One then finds out that the universal  
wave function does not depend on time\footnote{In some non-standard proposals,  
like, e.g., in Ref.~\ref{Valentini}(b), the universal wave function {\em  
does} depend on time and does not respect the Wheeler-De~Witt equation of  
the canonical quantum gravity.} (which is a well-known  
fact, see, e.g., our paper\ci{Shtanov} for treatment in the pilot wave  
formulation) so that the corresponding ``measure density'' {\em is}  
invariant. A subsystem of such a universe can happen to be sufficiently 
``disentangled'' from the rest 
of the world, at the same time exhibiting ergodic dynamics. 
Since the total wave  
function is time-independent, the wave function of such a ``disentangled''
subsystem will also be stationary, and the following    
reasoning will apply to this subsystem. 

 Consider, then, a system in a stationary ergodic state with a 
square integrable wave function $\Psi$. The ergodicity guarantees that  
the average time spent by the system in any region of its configuration 
variables is  
proportional to the measure of that region with measure density $|\Psi|^2$,  
as required. For the preparation process the ergodicity  
argument proceeds as follows.  Let $z = (x, \, y)$ denote the    
configuration variables of the total system, where $x$ represents the 
coordinates of the  
subsystem of interest, and $y$ the coordinates of the environment.  Let the  
total wave function $\Psi (z)$ have a structure  
\beq  
\Psi (z) = \psi(x) \phi (y) + \Psi_0 (z) \, , \label{psiuni}  
\eeq  
in which $\phi (y)$ is nonvanishing in a region $\Omega$ of the variables
$y$, which is  
complementary to the $y$-support of $\Psi_0 (z)$.  Then every  
time the corresponding piloted configuration variable $Y$ gets into the  
region $\Omega$, the configuration variable $X$ is piloted by the  
wave function $\psi (x)$. Provided $Y$ is in the region $\Omega$, the 
probability that $X$ will be in a region $\omega$ of the variables $x$ 
is given by the limit of the corresponding time ratio as   
\beq  
\p(X \in \omega \, | \, Y \in \Omega) = \lim_{T \to \infty} {\int_0^T  
\chi_{\omega \times \Omega} \left( Z (t) \right)\, d t \over \int_0^T 
\chi_{\C \times \Omega} \left( Z(t) \right)\, d t } \, , \label{p}  
\eeq 
where $\C$ is the whole domain of $x$, and $\chi_M$  
denotes the characteristic function of a set $M$. Since the evolution 
governed by $\Psi(z)$ is ergodic,\footnote{To avoid misunderstanding, note 
that the dynamics of $X$ when governed by $\psi(x)$ need not be ergodic.} 
then, according to the Birkhoff-Khinchin ergodic theorem  
(see Ref.~\ref{HWSLM}), the limit in Eq.~(\ref{p}) exists for almost every  
initial value of $Z$ and results in   
\beq 
\p(X \in \omega \, | \, Y \in \Omega) = {\mu_\Psi \left(\omega \times  
\Omega \right) \over \mu_\Psi \left( \C \times \Omega \right)}  
= \mu_\psi (\omega) \equiv \int \limits_\omega d \mu_\psi \, , \label{p1} 
\eeq 
where $\mu_\Psi$ and $\mu_\psi$ are the measures in the  
domains, respectively, of $z$ and of $x$ with densities determined by  
the corresponding normalised wave functions.  
The characteristics of the region $\Omega$ disappear 
from the result (\ref{p1}), and, if necessary, one can apply a formal limit 
of infinite-dimensional domain of $y$. The equality (\ref{p1}) in principle 
constitutes the justification of the standard quantum probabilities. 
 
 Of course, there are states that do not lead to ergodic evolution, like,  
e.g., a state with a real wave function. One should {\em assume} that the   
system of relevance is in a state that is close to  
ergodic. This is our specification of {\em complicated} systems, or, rather,  
complicated states.  The conditions under which the quantum evolution is  
ergodic must be further studied, just as it is the case with the classical 
ergodic theory of equilibrium. We leave this as a matter of future  
investigation. Some examples are presented in the last section. 

 At this point it is worth mentioning one interesting possibility of 
generating ergodic motions. Let 
\beq
\dot X = v(X) \, \label{guide}
\eeq
be the guidance equation in the original approach of Bohm (as in 
Eq.~(\ref{guidcon}) for the nonrelativistic case) with the time-independent 
generalised velocity $v(X)$. To the right-hand side of Eq.~(\ref{guide}) one
can always add an arbitrary extra term $v^\prime(X)$, such that 
\beq
{\rm div}_X \left(\left|\psi(X)\right|^2 v^\prime(X) \right) = 0 \, .
\eeq 
Then the new guidance equation 
\beq
\dot X = v(X) + v^\prime(X)\, \label{newguide}
\eeq
will still define a measure-preserving flow. Presumably the 
velocity $v^\prime(X)$ can be chosen so complicated that the flow 
(\ref{newguide}) will be (close to) ergodic, and, at the same time the effect 
of $v^\prime(x)$ will be unobservable on
macroscopic scale. The modification (\ref{newguide}) is well in 
the spirit of the proposals of Refs.~\ref{BV} and \ref{Nelson}, the difference is that 
in our case the extra velocity term $v^\prime(X)$ is not of stochastic nature. 
 
 Perhaps, comments are required concerning the nature of the time parameter  
$t$ in Eq.~(\ref{p}). 
This parameter is associated with the time translation 
symmetry of the quantum dynamics of our closed system.   
With respect to this time parameter the evolution operator 
$U_t: \C_z \rightarrow \C_z$ that acts in the configuration space 
$\C_z$ of our system, forms a  
one-parameter group: $U_{t+s} = U_t \, U_s$. The integration measure  
in Eq.~(\ref{p}) is the Lebesgue measure on the real axis of $t$. This is  
the only measure that we must take for granted in the present approach.  
Mathematically, this measure arises due to the ergodic theorem. Physically,  
it reflects the fact that experiment with the $x$ system in the state  
$\psi(x)$ will start at a random moment of time with uniform probability 
distribution, the only natural probability distribution in the context of  
stationarity. 

 One can possibly improve the above argument in several ways. One of them 
is to consider $N$ identical systems described by the  
corresponding collections of variables $x_1, \ldots, x_N$. The total wave 
function will be a function of these variables, as well
as of the environment variables $y$. Let then $z = (x_1, \ldots, x_N, y)$ 
be the whole set of configuration variables, and let the whole range of $z$ 
contain $N$ regions $\Theta_1, \ldots, \Theta_N$ that may overlap, 
with the following 
properties: when $z \in \Theta_n$ then the wave function acquires an 
approximate form
\beq
\Psi (z) \approx \psi(x_n) \phi (x_1, \ldots, \hat x_n, 
\ldots, x_N, y)\, , \label{psiuni1}
\eeq
for the values of $x_n$ that form a set of measure $\mu_\psi$ close to unity, 
so that the 
$n$th system is guided by the wave function of the form $\psi(x)$.
Then one will be interested in the probability that, provided $z$ is in one of 
the regions $\Theta_n$, the corresponding variable $x_n$ is in a certain 
region $\omega$ in the configuration space of $x$. 
This probability, according to the ergodic theory, will 
again be given by an appropriate time ratio (which one can easily write down) 
similar to Eq.~(\ref{p}), what 
will result in the last expression of Eq.~(\ref{p1}).  In this case, however, 
an observer will have $N$ systems at his disposal, and a large region 
$\Theta_1 \cup \Theta_2 \cup \cdots \cup \Theta_N$ of ``recurrence,'' 
so that equilibrium time 
average in the ensemble of $N$ systems will be achieved more rapidly  
as compared to the case of only one such system. Note that the situation just 
described is analogous to that of real experiments, in which ensembles 
are usually constituted of many different identical systems.  

 Next, if a system of interest, which is described by the coordinates $x$,
is part of a {\em large} closed ergodic system, then it follows that its 
equilibrium properties will be revealed only on long timescale, namely, on
the recurrence timescale of the whole system with respect to the region 
$\Omega$ described above. Therefore one needs ergodicity to take place 
on a sufficiently small scale. For instance, it may turn 
out that the coordinates $y$ of the environment can be partitioned in $M$ 
different ways into $y_m^\prime$ and $y_m^{\prime\prime}$, $m = 1,\ldots,M$ 
with the following additional property. When 
$y_m^{\prime\prime}$ is in a certain region $\Omega_m^{\prime\prime}$ 
the wave function (\ref{psiuni}) acquires the form 
\beq
\Psi (z) = \psi_m (x, y_m^\prime)\phi_m (y_m^{\prime\prime})\, , 
\label{psiuni2}
\eeq
such that the $(x, y_m^\prime)$ system is piloted by the wave function 
$\psi_m(x, y_m^\prime)$ that itself leads to an ergodic motion. 
Now, it might turn out that $\left(\C_m^\prime\times\Omega_m^{\prime\prime} 
\right)\, \cap\, \Omega \ne \emptyset$ ($\Omega$ being the region described 
after 
Eq.~(\ref{psiuni}) and $\C_m^\prime$ is the whole domain of $y_m^\prime$) 
for all or, at least, for several $m$. In this case 
the ergodicity argument will apply, at a time, 
to one of the $M$ subsystems described 
by the coordinates $(x, y_m^\prime)$, with smaller recurrence time.
In such a way an hierarchy of ergodic motions might take place, 
resulting in a sufficiently small equilibrium time for the $x$ system.   
It seems that the conditions of such a kind are likely to occur in nature.

\section{Ergodicity argument for a generic case}

 It is clear that from the ergodic point of view it is not so much 
stationarity that is important, as the property that a system of interest
acquires a specified wave function $\psi(x)$ repeatedly. Then 
whenever it is in the state $\psi(x)$ one can apply time averages to 
its various dynamical variables, and use ergodicity arguments to explain the
origin of quantum equilibrium distribution $p(x) = |\psi(x)|^2$. In this 
section we briefly discuss this more general case. 

 Consider a subsystem with configuration variables $x$ in an environment
with configuration variables $y$. We do not assume the total system to be 
in a stationary state. We, however, suppose that in interaction with the 
environment the $x$ system preserves its identity, and from time to time 
acquires a specific wave function $\psi(x)$, what means that the total 
wave function $\Psi(x,y)$ factorises as $\psi(x)\phi_n(y)$ at moments of 
time $t_n$, $n = \ldots, -1,0,1,2,\ldots$\@ Then it may occur that the 
dynamics of the 
configuration variables $Y$ almost does not depend on that of $X$, while 
influencing strongly the behaviour of $X$. For instance, $Y$ may be 
semiclassical variables of very massive objects. This property may take place 
at least on certain timescale (of order $t_n - t_{n-1}$) large as compared 
to the timescale of motion of $X$, what will be sufficient for the following 
argument. As an example, the reader may have 
in mind a gas of molecules, with $\psi(x)$ describing the internal state
(electronic configuration) of all the molecules, and $y$ corresponding to 
their centre-of-mass coordinates. The $x$ system will be perturbed from 
time to time by the influence from the dynamics of $Y$. As before, let 
$\C$ be the space of configuration variables $x$. Then a sequence 
of maps $U_n: \C \rightarrow \C$ will emerge, such that $U_n(x) = X(t_n,x)$, 
where $X(t_n,x)$ is the configuration of $X$ at the moment $t_n$ provided its 
configuration at the moment $t_{n-1}$ is $x$. Due to manifestly chaotic
influence from the dynamics of $Y$ the maps $U_n$ in totality 
are likely to form an ergodic sequence of transformations. This will lead to 
establishment of quantum equilibrium for the $x$ system in the sense of 
time averages, as discussed above. Thus processes that occur on relatively 
large mass scales can influence the small-scale quantum dynamics leading to 
ergodicity of the latter.

\section{Examples} 
 
 The following examples will illustrate our approach.  
Consider a free quantum particle of mass $m$ on a torus  
$T^n = S^1 \times \cdots \times S^1$ ($n$ times)  
with period lengths $l_1, \ldots, l_n$. Stationary states with definite  
momenta are in a standard correspondence with sequences of integers \{$n_i$,  
$i = 1, \ldots,  n$\}. If the squared lengths $\{l_i^2\}$ are rationally 
independent, then the pilot wave dynamic flow $x_i (t) = x_i (0) + v_i t\,  
({\rm mod}\, l_i)$ with $v_i = 2\pi\hbar n_i / m l_i$, $i = 1, \ldots, n$  
will be ergodic for integers $\{n_i\}$ all different from zero. The invariant  
measure in these states is just the uniform Lebesgue measure, and from  
the ergodicity argument it follows that the average  
time spent by the particle in a region $\Omega$ will be proportional to the 
invariant measure of that region.  The situation becomes very simple in the  
particular case of a circle $S^1$. That in a stationary state with   
nonzero momentum the mean time spent by the particle in any segment is  
proportional to the integral of $|\psi(x)|^2$ over that segment, is easily  
verified without recourse to the ergodic theory. Note that in the above  
example the approach of Valentini\ci{Valentini} would achieve no goal, since any  
probability distribution $p(x)$ would be simply translated along the torus  
retaining its shape. From the viewpoint expressed in Ref.~\ref{Valentini} the 
systems just considered would be regarded as not sufficiently complicated.

 As a realistic example, consider molecular collisions in a gas. Before two 
molecules approach each other their electrons behave independently, piloted
by intrinsic wave function $\psi(x)$. During a collision the electronic
motions become perturbed by interaction, but after the molecules fly apart 
the electrons inside each of them are again piloted by the old wave function 
$\psi(x)$, if the probability of electronic excitation is low (what we assume).
Suppose that in the process of collision the motion of molecular centres of 
mass are to a sufficiently large extent independent of their electrons' 
motions. Then the process of collision induces a map $U: \C_1 \times \C_2 
\rightarrow \C_1 \times \C_2$, where $\C_1$ and $\C_2$ are, respectively, 
the domains of the electrons' configurations $x_1$ (of the first molecule) 
and $x_2$ (of the second molecule). This map preserves the measure with 
density $|\psi(x_1)\psi(x_2)|^2$ and the sequence of such maps due to 
repeatable collisions is likely to be ergodic. Due to this ergodicity an 
equilibrium distribution of the electrons' configurations in molecules will 
be established. 
  
 It is interesting to note that certain cases appear to be tractable in a 
rather simple way, and do not 
require quantum equilibrium for {\em all} the configuration variables prior 
to the experiment. For example, consider a simplified model (see, e.g., 
Bohm\ci{BohmQM}) of the Stern-Gerlach 
experiment in which a measurement of the electrons' total spin in 
$z$-direction is performed in an atom like silver, with zero orbital angular 
momentum and total electrons' spin equal to $\hbar / 2$.  
Let $q$ denote the coordinates of the centre of mass of the  
atom, and $x$ the coordinates of the electrons with respect 
to the centre of mass. 
The wave function has one free spinor index, and can be presented as 
$\sum_\alpha c_\alpha\Psi_\alpha (q,x)\ket\alpha$, 
with $\alpha \in \left\{\uparrow, \downarrow\right\}$ describing the 
$z$-component of the electrons' spin, and $c_\alpha$ being constants
such that $\sum_\alpha\left|c_\alpha\right|^2 = 1$. 
The Hamiltonian that describes the experiment contains the 
part $H_{\rm int}$ that describes interaction of the system with the 
magnetic field of the Stern-Gerlach apparatus. Since the electrons' wave 
function is localised within a tiny region around the nucleus, 
in solving the Schr\"odinger equation we can 
approximate $H_{\rm int}$ by $\sigma_z V_q$, where $V_q$ acts only on the 
centre of mass coordinates $q$, and $\sigma_z$ is the Pauli matrix that 
describes the electrons' spin, so that the total Hamiltonian is 
\beq
H = H_x + H_q + \sigma_z V_q \, , \label{sgham}
\eeq
where $H_x$ and $H_q$ are the corresponding free Hamiltonians for 
$x$ and $q$. Let the initial wave function be 
\beq
\Psi(q,x)\sum_\alpha c_\alpha\ket\alpha  = \phi(q)\psi(x) \sum_\alpha 
c_\alpha\ket\alpha \, , \label{wf}
\eeq
where $\phi(q)$ is a localised wave packet, $\sum_\alpha\psi(x)\ket\alpha$ 
is the electrons' state vector, and
\beq
H_x \psi\ket\alpha = E \psi\ket\alpha \, . 
\eeq
The evolution will be governed by the Schr\"odinger equation with the
Hamiltonian of Eq.~(\ref{sgham}), and the solution for the wave function 
can be presented as 
\beq
\sum_\alpha c_\alpha\Psi_\alpha(q,x,t)\ket\alpha = 
\psi(x)\exp\left(-i{E \over \hbar}t\right)
\sum_\alpha c_\alpha\phi_\alpha(q,t)\ket\alpha \, , \label{sol}
\eeq
with $\phi_\alpha(q,t)$ being the solution to the Schr\"odinger equation 
\beq
i\hbar\dot\phi_\alpha = \left(H_q + s_\alpha V_q \right)
\phi_\alpha \, , \ \ \ s_\uparrow = 1 \, , \ \ s_\downarrow 
= - 1 \, , \label{phi}
\eeq
with the initial condition $\phi_\alpha(q,0) = \phi(q)$. 
It is easy to show that, in view of the factorisation in Eq.~(\ref{sol}),
the guidance equation for the configuration variables $Q$ that correspond
to $q$ is the same as what would 
stem from the equation (\ref{phi}), were the latter the genuine 
Schr\"odinger equation for a two-component spinor $\phi_\alpha$.
Thus, in the approximation Eq.~(\ref{sgham}) considered,  
the evolution of $Q$ {\em does not depend} on the evolution of $X$, 
the configuration variables that 
correspond to $x$. The experiment must be arranged in such a way that the 
wave packets $\phi_\uparrow(q,t)$ and $\phi_\downarrow(q,t)$ become 
nonoverlapping, and the variables $Q$ will then enter one of them. The 
probabilities $p_\alpha$ of entering $\phi_\alpha(q,t)$, 
will be given by the standard expression, $p_\alpha = \left| c_\alpha
\right|^2$, provided the initial distribution $p(q)$ of the variables 
$Q$ is $p(q) = \left|\phi(q)\right|^2$. Remarkably, the probabilities 
$p_\alpha$ do not depend on the initial distribution of $X$.

\begin{figure}[ht]
\centerline{
\epsfig{figure=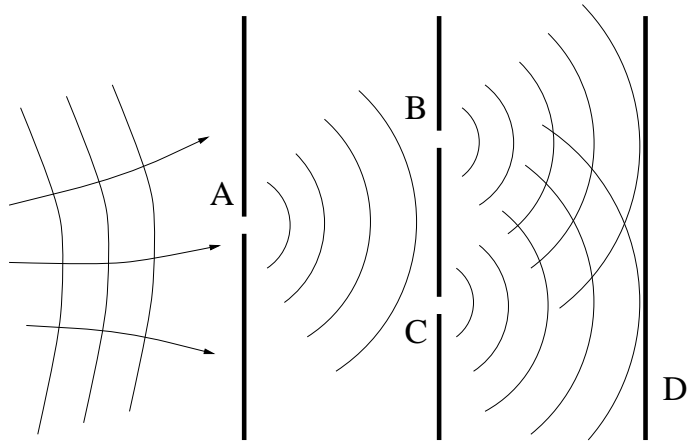,width=0.55\textwidth}}
\caption[ ]{Two-slit experiment} \label{fig1}
\end{figure}  

 As another example consider the classical two-slit experiment 
(see Fig.~\ref{fig1}). A system of  
collimating instruments and velocity selectors (not shown in the figure) to  
the left of the slit $A$ filters out particle wave functions, letting those  
with wavelengths in a sufficiently narrow band to pass to reach the slit $A$.  
The role of the slit $A$ is to produce spherical monochromatic waves in the  
space to the right of the slit $A$, this requires its dimensions to be much  
smaller than the wavelength of the wave function in the space to the left of  
the slit $A$. The spherical waves produced are then diffracted on a pair of 
slits $B$  
and $C$.  Now, the appearance of the familiar interference pattern on the  
screen $D$ will take place provided particles fall onto the slit $A$  
uniformly within the slit's range (since the wave function is also uniform on  
the scale of the slit dimensions, the condition $p = |\psi|^2$ will hold in  
the vicinity of the slit $A$, hence, by equivariance property it will hold  
also in the space to the right of $A$, in particular, on the screen $D$). But  
this last condition can be easily granted since the dimensions of the slit  
$A$ are small. Thus, whatever of {\em continuous} particle distributions is  
realised to the left of $A$ (its spatial scale of continuity is comparable to  
the spatial scale of particle wave functions), the familiar interference  
pattern will appear on the screen. This example suggests that the strong 
condition of quantum equilibrium achieved on the {\em universal}  
level may be not necessary. And, to support the agreement between the actual  
quantum experiments and the predictions of the pilot wave theory it is only  
necessary that quantum equilibrium distribution arises in {\em preparation}  
processes (natural or artificial).  For example, in the case of the two-slit  
experiment (Fig.~\ref{fig1}) quantum equilibrium distribution arises in the  
space to the right of the slit $A$, even though it may not take place to the  
left of $A$.

For completeness, we remark on the role of observer in 
quantum experiments, and in quantum theory in general. 
We are far from reducing a (human) observer to just a part of the universe 
subject to deterministic laws. Granted with a considerable 
freedom of action (one aspect of free will) he can influence the natural 
processes thus being able to make experiments. In a physical experiment the 
role of 
an observer is to put certain physical systems into contact so that they 
start to interact, while preventing some other systems from such a contact. 
In the course of a typical quantum-mechanical experiment an ensemble of 
identical systems will be filtered out and then made interact with parts 
of the experimental equipment. Such processes usually are described in 
terms of external influence by an experimental equipment on 
an ensemble of quantum systems. This description may involve time-dependent 
wave functions of quantum ensembles, and then the property of 
equivariance will preserve the quantum equilibrium condition 
$p(x) = |\psi(x)|^2$ for an ensemble under consideration. 
In nature interaction between 
different more or less sharply identified systems occurs by the law 
of chance (from human point of view), but is described similarly to 
the laboratory experiments.

\section*{Acknowledgments}

The author acknowledges kind hospitality and highly stimulating 
atmosphere of IUCAA, where this paper was completed.

\section*{References}
 
\begin{enumerate}

\item\label{Bohm} D.~Bohm, {\it Phys. Rev.} {\bf 85}, 166 (1952);  
   {\bf 85}, 180 (1952). 

\item\label{de Broglie} L.~de~Broglie, J.~Physique, 6e s\'{e}rie {\bf 8}, 225 
   (1927); \ L.~de~Broglie, {\it Une Tentative d'Interpr\'{e}tation Causale 
   et non 
   Lin\'{e}aire de la M\'{e}canique Ondulatoire: la Th\'{e}orie de la Double 
   Solution} (Gauthier-Villars, Paris, 1956). English translation: 
   (Elsevier, Amsterdam, 1960). 

\item\label{Bell} J.~S.~Bell, {\it Speakable and Unspeakable in Quantum 
               Mechanics} (Cambridge University Press, Cambridge, 1987).

\item\label{BH}  D.~Bohm and B.~J.~Hiley, {\it The Undivided Universe: An 
   Ontological Interpretation of Quantum Theory} (Routledge, 
   London and New York, 1993). 

\item\label{Holland} P.~Holland, {\it The Quantum Theory of Motion}  
   (Cambridge University Press, Cambridge, 1993). 

\item\label{Horiguchi} T.~Horiguchi, {\it Mod. Phys. Lett. A} {\bf 9}, 1429 
	(1994).  

\item\label{Shtanov} Yu.~V.~Shtanov, {\it Phys. Rev. D} {\bf 54}, 2564 (1996).

\item\label{Valentini} A.~Valentini, (a)~{\it Phys. Lett.} {\bf A156}, 5 (1991);  
   {\bf A158}, 1 (1991); \ (b)~{\it ``On the  
   Pilot-Wave Theory of Classical, Quantum and Subquantum Physics,''}  
   Ph.D. thesis, ISAS (Trieste, 1992). 

\item\label{DGZ} D.~D\"{u}rr, S.~Goldstein and N.~Zangh\`{\i}, 
	{\it J. Stat. Phys.} {\bf 67}, 843 (1992). 

\item\label{Bohm1} D.~Bohm, {\it Phys.Rev.} {\bf 89}, 458 (1953). 

\item\label{BV} D.~Bohm and J.-P.~Vigier, {\it Phys. Rev.} {\bf 96}, 208 (1954).

\item\label{Nelson} E.~Nelson, {\it Phys. Rev.} {\bf 150}, 1079 (1966). 

\item\label{LL} L.~D.~Landau and E.~M.~Lifshits, {\it Statistical Physics}  
   Part~I, 3d edition (Pergamon Press, 1994). 

\item\label{HWSLM} P.~R.~Halmos, {\it Lectures on Ergodic Theory} (The  
   Mathematical Society of Japan, Tokyo, 1956); \ P.~Walters,  
   {\it An Introduction to Ergodic Theory} (Springer-Verlag, 1981); \  
   Ya.~G.~Sinai, {\it Topics in Ergodic Theory} (Princeton University  
   Press, Princeton, New Jersey, 1994); \ 
   A.~Lasota and M.~C.~Mackey, {\it Chaos, Fractals and Noise: Stochastic
   Aspects of Dynamics}, Second Edition (Springer-Verlag, 1994). 

\item\label{BohmQM} D.~Bohm, {\it Quantum Theory} (Prentice-Hall,  
   N.Y., 1951). 

\end{enumerate}
\end{document}